\documentclass[aps,showpacs]{revtex4}
\usepackage{mathrsfs}
\usepackage{pifont}

\usepackage{graphicx}
\usepackage{amsmath,amssymb,amsthm}
\usepackage[colorlinks=true,urlcolor=blue,linkcolor=blue,citecolor=green]{hyperref}
\newcommand{\beq}{\begin{equation}}
\newcommand{\eeq}{\end{equation}}
\newcommand{\bea}{\begin{eqnarray}}
\newcommand{\eea}{\end{eqnarray}}

\begin{document}

\title{Type II see-saw dominance in $SO(10)$ }

\author{
Alejandra Melfo$^1$, Alba Ram\' irez$^1$ and  Goran Senjanovi\'c$^2$}

\affiliation{ $^1${\it Centro de F\'{\i}sica Fundamental,
Universidad de Los Andes, M\'erida, Venezuela}, $^2${\it International Centre for Theoretical Physics, 34100 Trieste, Italy } }
\begin{abstract}
 Grand unified theories where the neutrino mass is given by Type II seesaw have the potential to provide interesting connections between the neutrino and  charged fermion sectors. We explore the possibility of having  a dominant Type II seesaw contribution in supersymmetric SO(10). We show that this can be achieved in the model where symmetry breaking is triggered by $54$ and $45$ dimensional representations, without the need for additional fields other than those already required to have a realistic charged fermion mass spectrum.  Physical consequences, such as the implementation of the BSV mechanism, the possibility of the fields responsible for Type II see-saw dominance being  messengers of supersymmetry breaking, and the realization of baryo and leptogenesis in this theories are discussed.

\end{abstract}
\pacs{12.10.Dm,12.10.Kt,12.60.Jv}
\maketitle

\section{Introduction}

 The construction of a Grand Unified theory has reached new standards, after the minimal supersymmetric model based on  $SO(10)$ was proposed \cite{oldso10,plb}, and a precise calculation of the mass spectrum \cite{aarti,fukuyama,partone,charan} allowed for detailed fitting of fermion mass parameters   \cite{Matsuda:2000zp,Goh:2003sy,Babu:2005ia,Aulakh:2005mw,Bajc:2005qe,Bertolini:2005qb,Bertolini:2006pe}.

 The minimal  $SO(10)$ model, also known as MSGUT, has the smallest indispensable number of couplings in the superpotential as it uses only the strictly necessary representations.
  Fermion masses, including Majorana neutrino masses for the see-saw, are obtained through the vacuum expectation value (vev) of the ${\bf \overline{126}}$ representation,
  and an additional  ${\bf 126}$ keeps supersymmetry from being broken by D terms. A realistic fermion mass spectrum requires another Yukawa coupling, this  can be achieved by including the ${\bf 10}$ representation.   To make sure the electroweak Higgs is a linear combination of at least two doublets, a coupling between  ${\bf \overline{126}}$  and ${\bf 10}$  is required, and this is provided  by a ${\bf  210}$ field \cite{Babu:1992ia}.  These four Higgs fields are sufficient, for ${\bf 210}$ and ${\bf 126}$ can also achieve the symmetry breaking all the way down to the MSSM. In addition, as in any supersymmetric theory with a see-saw mechanism \cite{seesaw}, the R-parity symmetry which is in the center of the group  $SO(10)$ remains unbroken at low energies, providing proton stability \cite{rparity} and the lightest supersymmetric partner as a Dark Matter candidate. All these features require only 26 parameters in the superpotential, making the theory tantalizingly predictive.

 The fact that the theory is even too predictive for its own good is perhaps no surprise.  The superpotential has so few parameters that the composition of the light MSSM Higgses is
 completely determined by the symmetry breaking up to basically one relevant parameter, and so are the neutrino masses\cite{Bajc:2005qe}.  It is now generally accepted that the
  MSGUT  cannot account for large enough neutrino masses for the most general combination of Type I and Type II see-saw \cite{Aulakh:2005mw,Bajc:2005qe,Bertolini:2006pe}. In
   the case of Type II, this could have been suspected from a theory that does not allow for an intermediate scale below the unification scale $M_X$: roughly speaking, the left-handed triplet providing a direct Majorana mass for the left-handed neutrino, is too heavy.  In The Type I case, the problem is that  if ${\bf \overline{126}}$ is to play any role in the charged fermion masses, its Yukawa couplings cannot be arbitrarily small. The fitting requires them to be quite large, and since the Type I see-saw mass is proportional to the inverse Yukawa matrix, again it turns out too small. The current status is that this model is not viable in its minimal form (for reviews, see \cite{revgut}).

 One is tempted to accept the defeat and abandon the theory altogether, for any extension from the minimal model destroys predictivity. A possibility is to keep the minimal theory but not  to assume low energy supersymmetry. In fact, the theory can work \cite{Bajc:2008dc} with strongly split supersymmetry \cite{ArkaniHamed:2004fb}. Splitting supersymmetry on the other
 hand makes the model not so special, since one can also have a theory with ${\bf \overline{16}}$ \cite{Bajc:2004hr} instead of ${\bf \overline{126}}$,  with a radiative seesaw mechanism \cite{Witten:1979nr}.
 If one is willing to do away with supersymmetry altogether, one could use the fact that non-supersymmetric unification not only allows for intermediate scales that could be associated with a $B-L$ breaking scale, but in fact needs them. Non-supersymmetric  $SO(10)$ models are viable, although much less predictive that the minimal GUT, and some general remarks about its Yukawa sector can be made \cite{Bajc:2002iw,Bajc:2005zf}.  Yet another possibility is to keep the theory intact and hope that the supersymmetry thresholds can be large enough to make it work  \cite{Hall:1985dx}.

However, it may be worthwhile to set one's scruples aside and  study the consequences of enlarging the minimal model.
One road is to change the Yukawa sector, allowing for a complete one, i.e. including the only other possibility of a ${\bf 120}$ representation, as pursued for example in
 \cite{Aulakh:2005mw,aulakh120pletfull,aulakh120pletmodels,grimus}. In this case the ${\bf \overline{126}}$ Yukawa couplings can be lowered to enhance Type I seesaw masses, but one must resort to 
 large supersymmetry breaking  threshold corrections in order  to lower down type quark Yukawa couplings from their SM values. This in turn can lead to significant constraints on the
 usable  soft Susy parameters, implying for instance large trilinear  soft Susy breaking scalar couplings and a heavy third scalar generation (see last paper in
 \cite{aulakh120pletfull}).     Another way  is to relax the constraints on the MSSM Higgs composition by changing the non-Yukawa sector, as done for example in \cite{Goh:2004fy}.

If the Yukawa sector is taken to be complete, even a model with minimal choice  of the symmetry-breaking potential has a large number of fields at our disposal (with, admittedly, the corresponding loss in predictivity). In this paper, we explore the possibility of using this plethora of fields to achieve Type II see-saw dominance, by identifying a series of states that cancel out the triplet contribution to the RGE equations and determining whether they can have a low mass \footnote{As strong advocates of minimality ourselves, we hope that History will forgive us. }.

Neutrino masses arising mainly from a Type II term are attractive for a number of  reasons. First, as shown by Bajc, Senjanovi\'c and Vissani (BSV) \cite{Bajc:2002iw,Bajc:2004fj}, small quark mixing angles and large atmospheric neutrino mixing are naturally obtained, and connected to
$b-\tau$ unification at the GUT scale. This prediction  by itself is an  appealing enough  feature as to justify the study. But additionally, the triplet responsible for Type II, together with the fields required to cancel out its contribution to the RGE equations, can be an attractive messenger for supersymmetry breaking as proposed by Joaquim and Rossi \cite{Joaquim:2006uz}. This provides relations between the neutrino mass parameters, lepton flavour violation in the slepton sector, and the sparticle spectra. Another interesting role for a light triplet appears in leptogenesis \cite{Hambye:2003ka}.

  It is certainly worth investigating whether an $SO(10)$ GUT can accommodate such scenario.
   In the model with a GUT Higgs breaking triggered by a ${\bf 210}$, this has been done in  \cite{Goh:2004fy} by adding a  ${\bf 54}$ representation, whose only role is to provide a pair of light ${\bf 15}$,${\bf  \overline{15}}$  of SU(5), that contain the Type II triplet. A different approach is followed in \cite{Frigerio:2008ai,Calibbi:2009wk}, where instead of enlarging the Higgs sector one adds extra matter fermions  in an additional  ${\bf 10}$ representation. This ensures Type II dominance and the survival of complete SU(5)  representations to protect unification. 
   
  On the other hand, the   ${\bf 54}$  serves perfectly well by itself for GUT symmetry breaking, and is a natural alternative to  ${\bf 210}$ as long as there is also ${\bf 45}$ at the renormalizable level, as proposed in \cite{Aulakh:2000sn,Drees:2008tc}. Why would one trade one representation for two, increasing the number of free parameters? One important advantage is the smaller number of fields in  ${\bf 54}$ + ${\bf 45}$, which renders the theory less infrared slaved.
  However, a realistic charged fermion mass spectrum requires the presence of ${\bf \overline{126}}$  and ${\bf 10}$ Yukawas, and without ${\bf 120}$ these fields cannot mix. In this sense, the minimal realistic renormalizable model with ${\bf 54}$  and ${\bf 45}$ GUT Higgs contains a complete Yukawa sector. We shall focus on this version of $SO(10)$ unification.

We start in the next section by exploring the possibility of having a light triplet in a renormalizable $SO(10)$ theory.
 To orientate the discussion, we focus on the minimal model with ${\bf 210}$.
 We show that realization of the Type II dominance scenario is problematic: the minimal model has too few parameters, and any deviation from minimality worsens the already pressing problem of perturbativity. Yet, it is possible to have a light triplet by  adding a ${\bf 54}$ or completing the Yukawa sector with a ${\bf 120}$, and the triplet mass is in principle only bounded from below by perturbativity concerns.

In Section III, we turn to the model with ${\bf 54}$ and ${\bf 45}$, more economical in terms of contributions to the RGE above $M_{GUT}$.
We first present a complete study of the symmetry breaking and calculate the mass spectra.  Next, we show that in this case there are two different settings where the triplet can remain light.
We examine the physical consequences in Section III.
For one of the settings, the possibility of supersymmetry mediation is easily realized, preserving most of the predictions of Ref. \cite{Joaquim:2006uz}. Connection of $b-\tau$ unification with the atmospheric mixing angle as in  \cite{Bajc:2002iw} is also achievable.   We conclude that Type II dominance can be implemented in an $SO(10)$ model without a need to enlarge the Higgs sector in an {\it ad hoc} manner.

\section{General considerations}

  Intermediate scales are not allowed in supersymmetric grand unification, simply due to the fact that there is a very precise,  one-step unification with a supersymmetric scale of the order TeV from the RGE equations at two-loop order \cite{susyunif}. The set of three linear equations for the gauge coupling constants at one-loop,
  \beq
  \alpha_i(M_W)^{-1} = \alpha_U^{-1} - \frac{b_i}{2\pi } \ln\left( \frac{M_{GUT}}{M_W}\right)
  \label{RGE}
  \eeq
 does not admit a second solution, and naturally this cannot change much at the two-loop level.  However, it is still possible to have a light triplet while all the vevs lie at the unification scale, if the triplet's contribution to the beta functions at one loop is exactly cancelled by the contribution of other light particles. A  way to achieve this is to arrange for one complete GUT multiplet to be light, as attempted in \cite{Goh:2004fy}, where an additional ${\bf 54}$ representation is added on top of the MSGUT in order to ensure a breaking through the intermediate group $SU(5)$  that leaves a complete 15-dimensional $SU(5)$  representation, containing the required $SU(2)$  triplet, light.

 Another possibility is to arrange for  the masses   of the triplets and additional light particles to lie below $M_{GUT}$. Notice that  this does not have to be a very fine tuning, namely, a couple of orders of magnitude should suffice.  In order to find the "magic" fields, we implement an algorithm that looks for all possible combinations of states  in a given model that satisfy the conditions
\beq
\sum_i \delta b^i_1 - \delta b^i_2 = \sum_i \delta b^i_1 - \delta b^i_3 = 0
\label{lighttrip}
\eeq
where $\delta b^i_1$ is the contribution of the state $i$ to the one-loop $\beta$ function for gauge coupling $g_k$, and demand that one of these states be the Type II triplet. Let us put this to work in the minimal model first.

 The so-called minimal supersymmetric  $SO(10)$ GUT  is remarkably simple. In addition to the  $ {\bf 126}, {\bf  \overline{126}}$ needed for neutrino mass, and the ${\bf 10}$ required  to have a realistic fermion mass spectrum, only ${\bf 210}$ is necessary to perform all the symmetry breaking. The Higgs superpotential is then

 \beq
W_H = m_1 \, {\bf 210}^2
+ m_2 \, {\bf 126}\, \overline{\bf 126}  + \lambda_1 \, {\bf 210}^3 + \eta\, {\bf 126}\, \overline{\bf 126} \,  {\bf 45}
+ m_3 {\bf 10}^2 +  {\bf 210}(\alpha
{\bf 126}+ \bar\alpha \overline{\bf 126})  \, {\bf 10}
\label{superpot210}
\eeq

This model has been thoroughly studied \cite{aarti,fukuyama,partone,charan}. In the MSSM minimum,  the masses of fields in   ${\bf 126}\, \overline{\bf 126} $ and ${\bf 210}$ can be written in terms of an overall mass scale, the couplings $\lambda_1$ and $\eta$, and a dimensionless parameter $x$ related to $m_1$ and $m_2$. The light Higgs are a combination of doublets in ${\bf 10},  {\bf 126}$ and $ \overline{\bf 126} $,  completely determined by the symmetry breaking.

Demanding conditions (\ref{lighttrip}) to be satisfied, we have found only three possible candidate sets that include the left-handed triplet. In terms of their $SU(3)\times SU(2)\times U(1)$ quantum numbers, they are:

\begin{description}
\item[  {\bf [A] }:]  $(8,1,0) + (3, 1 \pm 1/3) + (1,3,\pm 1)  $ (with $\sum_i \delta b^i_k = -4 $)

\item [  {\bf [B] }:]  $(6,1,\pm 2/3) + (3, 2 \pm 1/6) + (1,3,\pm 1)  $ (with $\sum_i \delta b^i_k = -7 $)

\item[  {\bf [C] }:]  $(6,1,\pm 1/3) + (1, 2 \pm 1/2) + (1,3,\pm 1)  $ (with $\sum_i \delta b^i_k = -5 $)

\end{description}

The first set of fields [{\bf A}]  contains a color triplet with hypercharge $\pm 1/3$, which mediates proton decay and therefore cannot be allowed to be light.  The second set [{\bf B}] is a complete $SU(5)$  representation, the {\bf 15}.
 In this model, it is contained entirely in $ \overline{\bf 126}$.  In  the MSSM minimum the masses of these fields are given by \cite{partone}

\begin{eqnarray}
m_{1,3,\pm 1} &=&- 4\,\left(\eta/\lambda_1\right) m_1\, x \left( 4\,{x}^{2}-3\,x+1 \right)
  /  \left( x-1 \right) ^{2} \\
m_{6,1,\pm 2/3} &=&  -4 \left(\eta/\lambda_1\right) m_1\, \left( 1-3\,x \right)  /
\left( x-1 \right) \\
m_{3, 2 \pm 1/6} &=&   - 2\,  \left(\eta/\lambda_1\right) m_1\, \left( 4\,{x}^{3}- 6\,{x}^{2}+ 5\,x- 1
 \right)  /  \left( x-1 \right) ^{2}
\end{eqnarray}
It is clearly not possible to adjust the only free parameter available, $x$, so that all of them are light. It can only happen in the $SU(5)$ -preserving minimum which is of no interest to us.

A similar situation happens for the set [{\bf C}] : the sextet and triplet fields cannot be simultaneously light. In this case notice that the set is not formed by a complete representation of any $SO(10)$ subgroup.

However as argued in the introduction, the minimal model cannot fit the fermion masses in any case. Adding a ${\bf 120}$ the most general superpotential becomes
     \beq
 \Delta W =  m_4 {\bf 120}^2 +
 \alpha \, {\bf 210} \, {\bf 10}\,  {\bf 120} +  (\beta \,{\bf 126}+  \bar \beta  \,\overline{\bf 126} )  \, {\bf 210} \, {\bf 120}
 + \lambda_2   {\bf 210} \,   {\bf 120}^2 +   (\gamma \,{\bf 126}+  \bar \gamma  \,\overline{\bf 126} )  \, {\bf 210} \, {\bf 10}
  \eeq
The symmetry breaking pattern is unchanged, and so are the masses of the triplet and the color sextet of case  [{\bf B}]. The sextet of set  [{\bf C}] however now is a mixture of fields in ${\bf 120}$ and $\overline{\bf 126}, {\bf 126}$. The doublets are a copy of the MSSM Higgs which have contributions from  all the representations, so in principle one would have at hand enough parameters to tune the masses of all three fields in the set to any desired value.

The problem with this scenario is  that any addition of representations  on top of the minimal  model with ${\bf 210}$  makes perturbativity a concern.     Imposing that the Type II triplets and their companion fields are also light only worsens the situation.

 One may be better off including a ${\bf 54}$ instead of the ${\bf 120}$, as in  \cite{Goh:2004fy}. There the proposal is slightly different: in order to keep the ${\bf 15}$ of $SU(5)$  light (our case  [{\bf C}] ), the symmetry breaking is done in two stages. $SO(10)$ is broken at a high scale $10^{18}$ GeV  down to an SU(5).  At $M_{GUT}$ $SU(5)$  is broken and one ${\bf 15}$ is made light by choosing some of the effective potential parameters, so that it only receives a mass from higher dimensional terms. The Type II triplet can then have a mass of order $10^{14} $ GeV.  However notice that  in fact adding a ${\bf 54}$ gives enough additional parameters to tune the mass of the three fields in ${\bf 15}$ to any value, and the same can be said for fields in case [{\bf C}] . Then the  lower limit on the mass of the triplet is set only by requiring perturbativity near the GUT scale.

\begin{figure}
\begin{center}
\includegraphics[width=6.5cm]{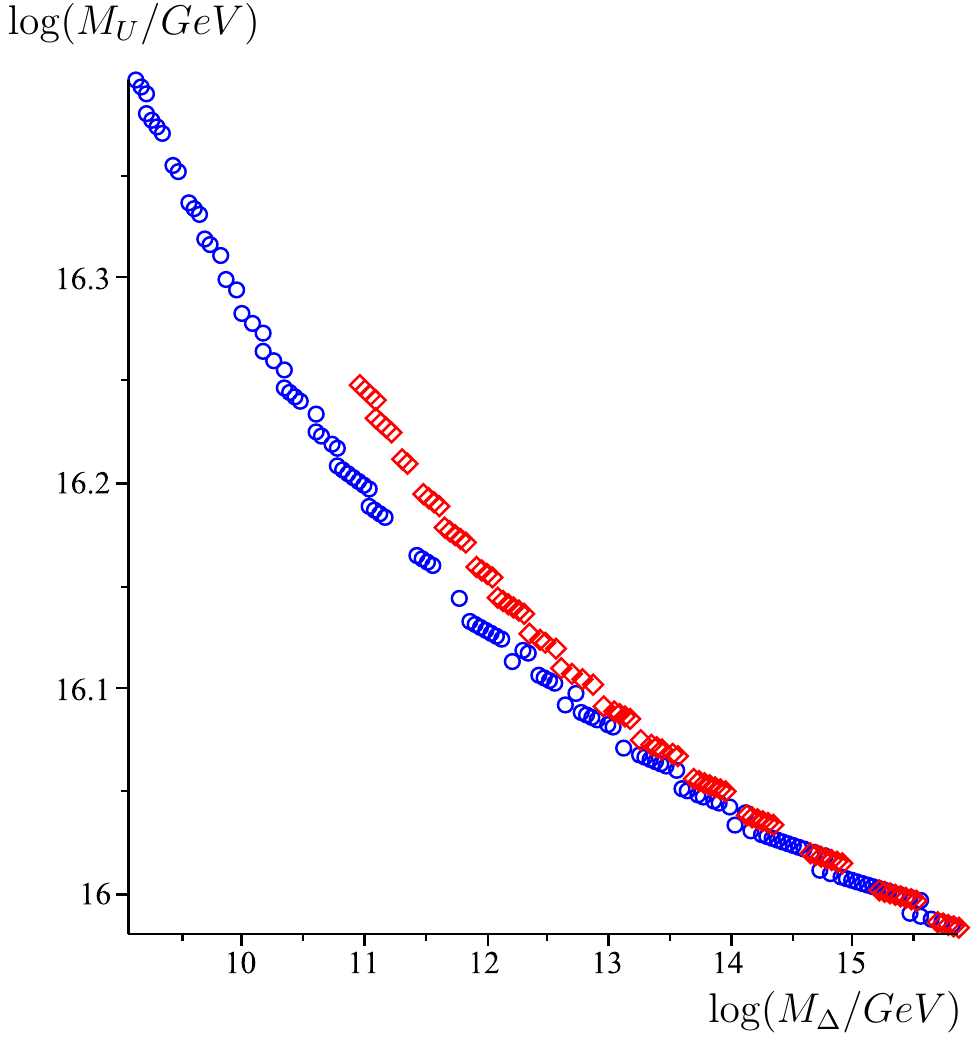}\includegraphics[width=6.5cm]{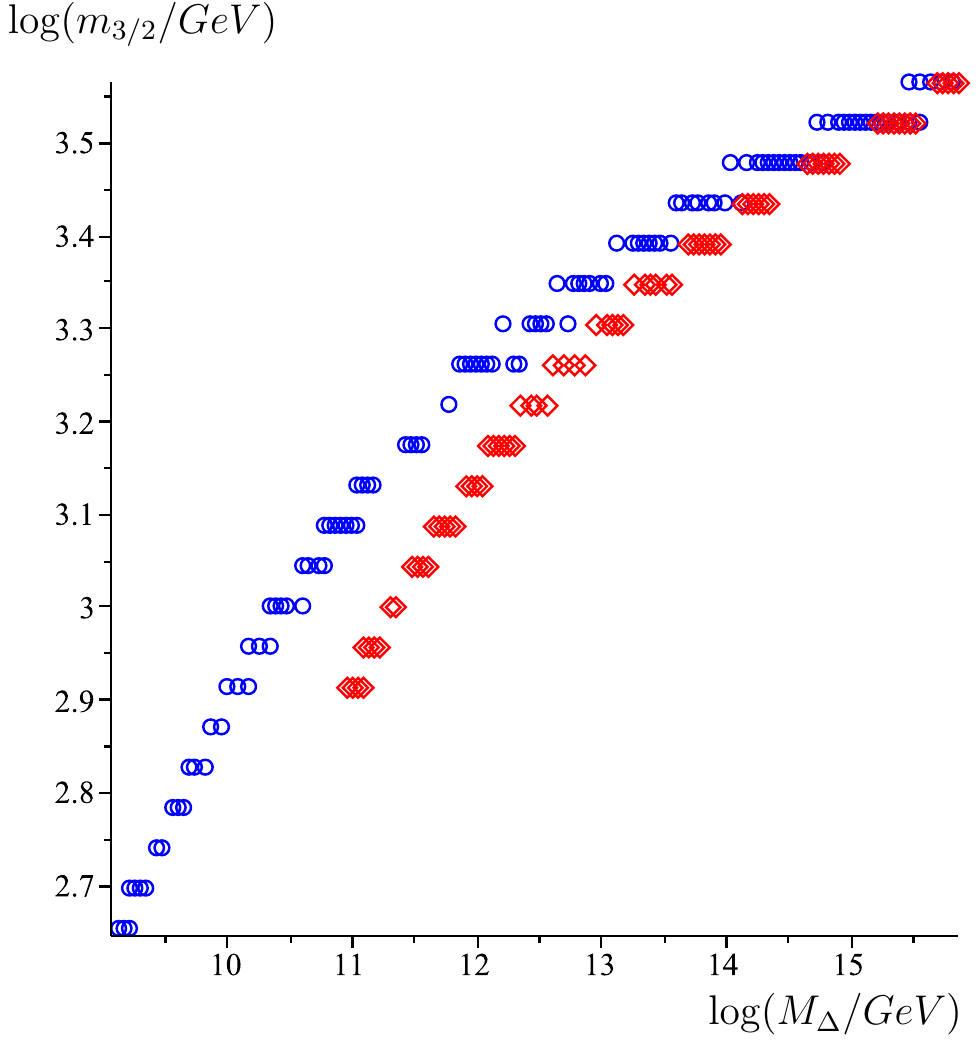} \\
\includegraphics[width=7cm]{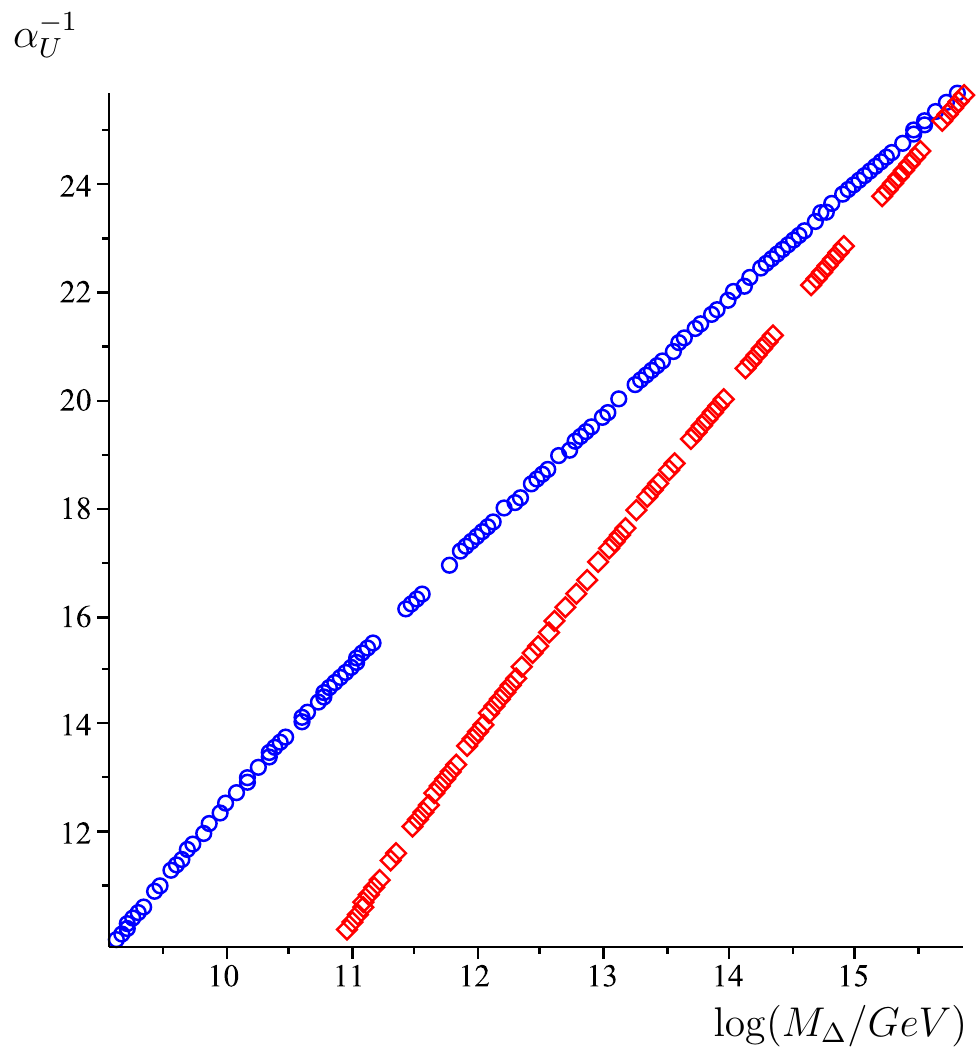}

\end{center}
\caption{Allowed values of supersymmetry  $\log(m_{3/2}/GeV)$, unification scale $\log(M_{U}/GeV)$ and  inverse unified gauge coupling $1/\alpha_U$  as a function of $\log(M_\Delta/GeV)$ for two-loop unification. $M_\Delta$ is the common mass scale of the left-handed triplet and the remaining fields in  sets   [{\bf B}]  (red diamonds) and   [{\bf C}] (blue circles).}
\label{f1}
\end{figure}

Conditions (\ref{lighttrip}) guarantee that the light fields do not spoil unification only at the one-loop level. And clearly, the unified coupling constant value is affected by the presence of the extra fields below $M_{GUT}$. Let us be more precise, by solving the two-loop equations for different values of the masses  of the fields in cases [{\bf B}] and [{\bf C}] ( keeping them equal).  Results are given in figure \ref{f1}.

  The effect on the unification scale is very small, it gets  raised for light triplets.  The  supersymmetric scale in turn gets slightly  lower. But the unified gauge coupling grows significantly.  Of course, threshold effects may be very important here. However it is useful to compare the contribution to the $\beta$ coefficients of all the fields in each of the models, as   in Table \ref{tablealfa}.
  \begin{table}[h]
 \begin{tabular}{l|c|c|c}
 Field content  & $b$   \\
 \hline
 ${\bf 10}$ + ${\bf 126}$ + ${\bf \overline{126}}$ +${\bf 210}$ & -109   \\
  ${\bf 10}$ + ${\bf 126}$ + ${\bf \overline{126}}$ +${\bf 210}$ + ${\bf 54}$ & -121   \\
  ${\bf 10}$ + ${\bf 126}$ + ${\bf \overline{126}}$ +${\bf 210}$ + ${\bf 120}$ & -137   \\
   ${\bf 10}$ + ${\bf 126}$ + ${\bf \overline{126}}$ + ${\bf 54}$ + ${\bf 45}$ + ${\bf 120}$& -101   \\
 \end{tabular}
 \caption{ Contribution to the $\beta$ function   in different  $SO(10)$ models}
 \label{tablealfa}
 \end{table}
We see that the most economical  version of the theory  is the one replacing ${\bf 210}$ by a ${\bf 54}$, ${\bf 45}$ pair. We now turn to discuss this model.

\section{SO(10) with ${\bf 54}$ and ${\bf 45}$}

\subsection{Symmetry breaking}

If ${\bf 210}$ fails to the the job, it is not difficult to pinpoint the alternative model.
Once we include  $ {\bf 126}, {\bf  \overline{126}}$, the smallest representation that can break the symmetry  is {\bf 45}. However, it is easy to  see that by itself it cannot achieve all the breaking:  the  $SU(5)$ singlets in {\bf 45} and {\bf 126},  ${\bf \overline{126}}$ take the vev, and the breaking stops there. The next possibility is the {\bf 54} representation, but again it cannot work by itself, for the singlet in {\bf 54} does not couple to the singlets in {\bf 126},  ${\bf \overline{126}}$. Both representations are therefore needed, as has been argued in  \cite{Aulakh:2000sn}, and are sufficient also.

In its turn, the  choice of these fields automatically implies that the Yukawa sector has to be complete, in the sense that all the representations that can couple to the fermions  have to be included.  The argument is as follows.  The fermions in $SO(10)$ are in the {\bf 16} dimensional spinor representation, and we have
\beq
{\bf 16} \times {\bf 16} = {\bf 10} + {\bf 120} + {\bf \overline{126}}
\eeq
A  {\bf$ \overline{126}$} field is indispensable for neutrino mass. It is then obvious that one cannot have just one Yukawa coupling, for the down quarks and charged leptons mass would be equal at $M_{GUT}$, for all generations. One needs at least one more coupling,  but this does not suffice, since it is necessary to ensure that the new field also acquires a vev. Or in other words, the MSSM Higgs doublets have to be a combination of doublets in at least two different representations. In the theory with {\bf 54 } and {\bf 45} the candidate is {\bf 120}, since it has a coupling
\beq
{\bf 45} \,  {\bf \overline{126}} \,  {\bf 120}
\eeq
 However, not even this suffices. As discussed in \cite{Bajc:2005zf}, the Yukawa sector with  ${\bf \overline{126}}$  and {\bf 120} only gives  wrong relations between the tau and the bottom quark  at $M_{GUT}$
 \beq
 m_\tau \sim 3 m_b
 \eeq
 while for a supersymmetric model, one would need $ m_\tau \sim m_b$ at the unification scale.
 The correct relations are instead provided by the combination of  ${\bf \overline{126}}$  and {\bf 10}, but neither {\bf 45} nor {\bf 54} can mix these doublets. They can only mix through the existence of {\bf 120}, for then one has the couplings
 \beq
 \alpha \, {\bf 45} \, {\bf 10} \,  {\bf 120} + \overline\beta \, {\bf 45} \, {\bf \overline{126}}\,  {\bf 120}
 \eeq
 To summarize, the Higgs sector is composed of
 \begin{itemize}
 \item  ${\bf \overline{126}}$, for neutrino mass, and of course  {\bf 126}
 \item {\bf 45} and {\bf 54} to help break the symmetry down to the MSSM
 \item {\bf 10} to have a realistic fermion mass spectrum
 \item {\bf 120} to provide the mixing  between the Higgs doublets
 \end{itemize}
  The most general superpotential  for the Higgs sector  can be written schematically as:
 \beq
 W =  \sum_{i=1}^4 m_i  \; \phi_i^2 + m_5  {\bf 126}\, \overline{\bf 126}+    \sum_{i=1}^6 \lambda_i \, \phi_i^2 \, {\bf 54}   + \eta\, {\bf 126}\, \overline{\bf 126} \,  {\bf 45}  +
 \alpha \, {\bf 45} \, {\bf 10}\,  {\bf 120} +  (\beta \,{\bf 126}+  \bar \beta  \,\overline{\bf 126} )  \, {\bf 45} \, {\bf 120}
  \eeq
where $i= (1,2,3,4,5,6) = ( {\bf 45} , {\bf 10},  {\bf 120},  {\bf 54} , {\bf 126}, \overline{\bf 126}  )$ .
The detailed superpotential is presented in the Appendix.

 As usual, it is convenient to decompose the  $SO(10)$ fields under the Pati-Salam $SU(4)_C\times SU(2)_L\times SU(2)_R$ group. The MSSM singlets in each representation are
   \begin{equation}
 \langle (1,1,1)_{\bf 54}\rangle \equiv  s  \;   ;    \;  \;  \langle (15,1,1)_{\bf 45}\rangle \equiv   a    \;   ; \;   \;
  \langle (1,1, 3)_{\bf 45}\rangle \equiv  b  \;   ; \;   \;
 \langle (10,1,3)_{\bf 126} \rangle \equiv  \sigma  \;   ; \;   \; \langle (\overline{10}, 1, 3)_{\bf \overline{126}} \rangle  \equiv  \overline{\sigma}
 \end{equation}
   So that the superpotential for these singlet fields is
     \begin{eqnarray}
W= 15 \,m_4\,s^{2}+20\,\lambda_4\,{s}^3  - \frac{m_1}{2}\,(3{a}^2+2{b}^2)  - 12\,s\,\lambda_1({a}^{2}-{b}^{2})
 +m_5\,\sigma\,\overline{\sigma}+\eta\,\sigma\,\overline{\sigma}\,(3a-2b)\nonumber
 \end{eqnarray}

The symmetry breaking equations
  \begin{eqnarray}
F_{s} &=& m_4 \,s +2 \lambda_{4} s^{2}-{2\over5}\lambda_{1}(a^{2}-b^{2})=0
\nonumber \\
 F_{a} &=&  m_{1}\, a  + 8  \lambda_{1}\, a\, s  - \eta\,\sigma \overline{\sigma}=0
\nonumber \\
 F_{b} &=& m_{1}\,b - 12 \lambda_{1}\, b\, s + \,\eta\,\sigma\, \overline{\sigma}=0
\nonumber \\
F_{\sigma} &=& m_{5}\, \overline{\sigma}+ \eta\,\overline{\sigma} (3a - 2b)=0
\nonumber \\
 F_{\overline{\sigma}} &=& m_{5}\,\sigma+ \eta\,\sigma (3a - 2b)=0
 \end{eqnarray}
have six different solutions presented in Table \ref{t1}. Five of them are either trivial or break the symmetry to intermediate groups, and are presented in Table \ref{t1}. The sixth is the one of interest, and is given by
 \begin{eqnarray}
  s&=&{m_{1}\over{4\lambda_{1}}}( 1- x) \nonumber \\
  a&=&{m_{5}\over{\eta}}{(2-3x)\over{5x}}  \nonumber \\
  b&=&{m_{5}\over{\eta}}{(3-2x)\over{5x}}  \nonumber \\
  \sigma&=&\sqrt{{m_{1}m_{5}\over{{\eta}^{2}}}{{(2- 3x)(3-2x)}\over{5 x}}}\nonumber \\
   \label{thevevs}
 \end{eqnarray}
  with
 \beq
{m_4\over{m_1}}=  {\left(x-1\right){\lambda_4\over{2\lambda_1}}} - \left({4\over5}{\lambda_1\over\eta}{m_5\over{m_1}}\right)^2{\left(x+1\right)\over{2x^2}}
\label{eqmassrel}
 \eeq
  The symmetry breaking is then completely specified by
 one overall mass scale,
  one mass ratio,
  one free complex parameter $x$
and the three couplings $\lambda_4$, $\lambda_1$, $\eta$.

\begin{table}
 \begin{tabular}{l|c|c|c|c}
 H & $s$ & $a$ & $b$ & $\sigma^2$ \\  \hline
  $SO(10)$ & 0 & 0 & 0 & 0 \\  \hline
$SU(4)\times SU(2)\times SU(2)$ & $-{m_4/{2\lambda_4}}$ & 0 & 0 & 0 \\   \hline
$SU(3)\times SU(2)\times SU(2)\times U(1)$&$-{m_1/{8\lambda_1}}$&$\pm \sqrt{5 m_1(  \lambda_4 m_1 - 4 \lambda_1 m_4)}/8\lambda_1^{3/2}$& 0&0 \\  \hline
$SU(3)\times SU(2)\times U(1)\times U(1)$ &$ {m_1/{12\lambda_1}}$& 0 &$\pm\sqrt{-5m_1(\lambda_4 m_1 + 6 \lambda_1 m_4)}/12\lambda_1^{3/2}$& 0  \\  \hline
$SU(5)$& 0 &$-{m_5 / {5\eta}}$&$ m_5 / 5\eta$ &$ - m_1 m_5/ 5 \eta^2$  \\
 \hline
\end{tabular}
\caption{Patterns of   incomplete symmetry breaking, $SO(10)\to H$. The sixth solution, corresponding to breaking to the MSSM, is discussed in the text. }
\label{t1}
\end{table}
The parameter $x$ can in principle be chosen arbitrarily, as long as particular points where the symmetry breaking is incomplete (such as $x=0, 1, 3/2$ or $2/3$) are avoided.  In order to be more precise, it is necessary to calculate explicitly the mass spectrum. This has been done in \cite{fukuyama} without specializing to a symmetry breaking pattern, by using the tensor methods of reference \cite{He:1990jw}.
 We have checked their results by calculating all masses directly: we first identify all Standard Model states in each of the  $SO(10)$ representations, and then compute the complete superpotential in terms of these states and the vevs. Then fermion mass matrices of states
 $(\psi_i, \bar\psi_i)$  are found by
 straightforward derivation
 \beq
 M_{ij} = \left[ \frac{\partial^2 W }{\partial\bar\psi_i \partial\psi_j } \right]_{\phi_i=0}
 \eeq
 We have found complete agreement with  \cite{fukuyama} up to overall  phase factors. Also, the matrices have been checked to provide the correct number of pseudo-Goldstone bosons in each of the symmetry breaking patterns, and in the case of  partial symmetry breaking, to give correct relations for states in the same multiplet.
The resulting mass pattern is shown in the Appendix.

\subsection{Arranging Type II see-saw dominance  }

 The possible sets of "magic"  fields in this model are found to be the same as in the ${\bf 210}$ version. But in this case the number of parameters is enough to tune all their masses as we now show.

 We first consider the complete 15-dimensional $SU(5)$ representation of set [{\bf B}].
 The mass matrices of $(6,1,\pm 2/3)$ and $(1,3,\pm 1)  $ ( numbers 1 and 6 in the Appendix respectively) will have a zero eigenvalue if
\beq
\frac{m_5}{m_1} = \frac{5 \eta x^2}{4 \lambda_1}\sqrt{\frac{\lambda_4(1 - 9x)}{2\lambda_1(1+x)}}
 \eeq
 with either
 \beq
 x=1 ; \,  {\rm or} \;  x =  \frac{13 R +2 \pm 5 \sqrt{R^2 + 4 R +1}}{12 R - 7} , \; R\equiv \frac{\lambda_1\lambda_5\lambda_6}{\lambda_4 \eta^2}
 \eeq
 The solution with $x=1$ means the symmetry is broken just up to $SU(5)$. However, the second solution will not give an $SU(5)$ symmetric vacuum as can be checked explicitly.
  Then, since the matrix (number 3 in the Appendix)  for the $(3, 2 \pm 1/6)$  states already contains a zero eigenvalue (the pseudo-Goldstone boson),   the equation
 \beq
 \sum_i^5 SDet_i(M_3) =0
 \eeq
 where $SDet_i$ are the minors of the $(i,i)$ element of the mass matrix, must be solved for, say, $m_3$. We do not include the solution due to its length. Once $m_5$, $m_3 $ and $x$ are fixed, $m_2$ is given by the requirement of fine-tuning of the light Higgs mass (matrix 17 on the Appendix). Thus the complete spectrum can be obtained in terms of one overall scale, $m_1$, and the couplings.  For example, one point is found to be
 \begin{eqnarray}
\frac{m_5}{m_1} &=& 0.9482892551,  \; \; \frac{m_3}{m_1} = 6.975439986 ,   \; \;  \frac{m_2}{m_1} = 0.2900315456,
\nonumber \\
  x&=& 1.384232261 ,  \; \;  \lambda_1=\lambda_2=-1,  \; \;  \eta=0.25 ,   \; \;  {\rm other \, couplings}=1
 \end{eqnarray}

  The third set of fields [{\bf C}] is more difficult to deal with, since it includes the doublet states and therefore contains the light SM Higgs.  The states do not  form a multiplet of any  $SO(10)$ subgroup, cancellation of their contributions to the $\beta$ coefficients is fortuitous. The corresponding matrices are number 6, 12 and 17 in the Appendix. Only matrix 6 can be solved independently for $m_5$, the other 3 zero eigenvalues have to be found by solving simultaneously the equations
\beq
Det(M_{12})=0 , \qquad  Det(M_{17})=0  , \qquad  \sum_i^5 SDet_i(M_{17}) =0
\eeq
for $m_3,m_2 $ and $x$. Equations are however intractable analytically. It is more convenient to set all couplings to a fixed value, and then solve numerically. One example point we have found is
 \begin{eqnarray}
\frac{m_5}{m_1} &=& 1.139049181,  \; \; \frac{m_3}{m_1} =-0.6168590170 ,   \; \;  \frac{m_2}{m_1} = -0.6736019338,
\nonumber \\
  x&=& 0.75 ,   \; \;  \alpha=1.129419783 ,   \; \;  {\rm other \, couplings}=1
 \end{eqnarray}
 Clearly, variations around this set of parameters can provide non-zero but equal values for the magic fields's masses.

 \section{Physical implications}

 \subsection{Supersymmetry breaking mediation}

In case  [{\bf B}] , one can implement the scenario of Joaquim and Rossi  \cite{Joaquim:2006uz} in a straightforward manner. Let us recall briefly how it works.
The essential point is that the Higgs triplet responsible for the type II seesaw acts at the same time as a messenger of supersymmetry
breaking. This in turns implies that the off-diagonal slepton masses are determined by the same Yukawa couplings that give the neutrino
mass matrix, and thus the leptonic flavor violation is controlled by the leptonic mixing matrix. In short, one ends up predicting the relative branching ratios of the muon and tau rare decays. Their work is based on the $SU(5)$  grand unified theory, with the addition of a full
$ \overline{\bf 15}$ multiplet. In their program the mass of this multiplet is not determined by the unification constraints, as opposed
to the situation discussed here. The advantage of $SO(10)$ is that this representation is already contained in the theory in order to correct
the mass relations of charged fermions, the price to pay is a more complicated setup . In any case, their main results regarding
LFV translate intactly.

The presence of the doublets in [{\bf C}]  poses a problem for the implementation of these fields as supersymmetry breaking messengers. All fields in $ \overline{\bf 126}$ would have to couple to the hidden sector, and while most of the fields in the representation would have masses at the GUT scale, it also takes part in the composition of the MSSM Higgs.
The MSSM Higgs as messengers of supersymmetry breaking is a rather interesting possibility as suggested in 
\cite{Dvali:1996hs}.
It has the problem of a tachyonic stop \cite{Dine:1996xk}, due to the large top Yukawa coupling. This could be in principle suppressed by a making the contribution of the $ \overline{\bf 126}$  to the light Higgs relatively smaller. A complete fit of fermion masses and mixings is then required, which
is beyond of the scope of the present work.

\subsection{Atmospheric neutrino mixing and $b-\tau$ unification }

Implementing the BSV mechanism of  \cite{Bajc:2002iw} is not so direct. Namely, now the Yukawa coupling with ${\bf120}$ also gives contributions to the fermion masses, and in a non-trivial way, since ${\bf 120}$ contains two different doublets: one is a singlet of $SU(4)$, (we shall denote it  $120_I$) , an the other is in the ${\bf 15}$  representation ($120_{II}$).
 Following the notation  of  \cite{Bajc:2004fj} one has the following expressions for the charged fermion mass matrices:
\begin{eqnarray}
M_U &=& v_{10}^u \, Y_{10} + v_{126}^u \, Y_{126} +   Y_{120} (v_{120_I}^u +   v_{120_{II}}^u ) \nonumber \\
M_D &=& v_{10}^d \, Y_{10} + v_{126}^d \, Y_{126} +   Y_{120} (v_{120_I}^d +   v_{120_{II}}^d ) \nonumber \\
M_E &=& v_{10}^d \, Y_{10}  - 3  v_{126}^d \, Y_{126} +   Y_{120} (v_{120_I}^d - 3   v_{120_{II}}^d )
\end{eqnarray}
while from Type II seesaw we have for the neutrino
\beq
M_\nu = Y_{126} v_T
\eeq
with $v_T$ the vev of the triplet. We have
\beq
M_\nu \propto M_D - M_E +  c \, Y_{120}
\eeq
with $c  $ a constant.

To see the effect of the ${\bf 120}$ couplings, let us consider the second and third generations only as in  \cite{Bajc:2002iw}. We work in the basis where $M_E$ is diagonal, so  $M_\nu$ determines the leptonic mixing angles.  We know the quark mixing angle is small, and it is plausible that the down quark mixing is small also (this was shown to be true in \cite{Bajc:2004fj} for the case without a ${\bf 120}$).

Therefore using the fact that the couplings of ${\bf 120}$ with fermions is asymmetric, neglecting second generation masses and down quark mixings we get
\beq
M_\nu \propto \left( \begin{array}{cc}  0 & c \\ -c & m_b-m_\tau\end{array}\right)
\eeq
This means that for $m_b \simeq m_\tau$, the atmospheric neutrino mixing angle $\theta_A \simeq 45^o$ for any value of $c$. Of course, this is a rough approximation. It may be that in the general case, in order for BSV to work $c$ is required to be small. This is a question worth of a more detailed analysis \cite{amrsnext}. We only point out here that $Y_{126}$ can in any case be made small,  since the  role of ${\bf 120}$ in this model is not to fix the fermion masses but to provide a mixing between ${\bf 10}$ and $\overline{\bf 126}$.

 \subsection{Baryo and leptogenesis, proton decay}

 It is rather tempting to try figure out, if possible, how baryogenesis  could have taken place in a theory such as the one discussed here.
 Unfortunately, it is not an easy task for there a number of different, perfectly realistic scenarios, that could lead to a baryon asymmetric Universe
 such as ours.

 {\it Electroweak baryogenesis}. It is well known that the MSSM is
 custom fit for the electroweak baryogenesis as long as  one of the stops is sufficiently light, lighter than the top quark \cite{Carena:1996wj}.
 The low energy-effective theory of the supersymmetric $SO(10)$ is precisely MSSM, so this would go intact if the stop was to be light.

{\it Affleck -Dine baryogenesis}.  It could turn out that the stops are heavy, so that  electro-weak baryogenesis does not work. That is not a
problem, for this theory has a number of flat directions which break baryon and/or lepton number. In particular, one has unbroken
R-parity which ensures such flat directions, offering a natural implementation of the mechanism of Affleck and Dine \cite{Affleck:1984fy} for baryogenesis. Even if the stop was light,
we would not know what caused the genesis, since both mechanisms could equally well be operative.

{\it Leptogenesis}.

 $SO(10)$ grand unified theory is an ideal framework for  leptogenesis \cite{Fukugita:1986hr} through right-handed neutrinos.
  This appealing scenario makes good use of the Majorana nature of right-handed neutrinos.
  The possible lightness of the Higgs triplet responsible for the Type II seesaw has important implications for this \cite{Hambye:2003ka}, in that
 it enlarges the parameter space. This may be important in a model restricted by the $SO(10)$
 symmetry, where neither the Yukawas nor the scales are arbitrary. A nice example  is the detailed calculation made by \cite{Frigerio:2008ai} in their model with additional matter fields, where constraints coming from requiring successful leptogenesis are later used to get constraints in flavour-violating observables \cite{Calibbi:2009wk}.

     However, the possibility of the co-existence of the above mechanisms for baryogenesis renders it virtually impossible to know in any near future, if ever,
      how genesis took place. Whether or not this is a curse of a blessing is up to the reader to decide.

{\it Proton decay}.

Grand unification was hoped to be originally the theory of proton decay, but the task of computing the decay rate and the
branching ratios is a tremendously hard task. In supersymmetric theories so much more since one does not know the
 spectrum of super partners. However, the d=5 operators in general lead to a very fast proton decay unless the GUT scale
is above $10^{17}\,$GeV or so \cite{Murayama:2001ur}. It turns out that the threshold effects in supersymmetric grand unified
theories are large enough even in the minimal SU(5) theory \cite{Bajc:2002pg} and GUT scale can be easily as large as
even the Planck scale. In other words, it is conceivable in principle that the proton decay be out of reach of the next generation
proton decay experiments now planned \cite{Senjanovic:2009kr}, which would could be quite a blow for grand unification. Fortunately, the consistency
of quantum gravity requires the upper limit \cite{Dvali:2007hz} on any scale of new physics $\Lambda \lesssim M_{Pl}/\sqrt {N(\Lambda)}$, where
$M_{Pl} \simeq 2 \times 10^{18}\,$GeV stands for the reduced Planck scale and $N(\Lambda)$ for the number of species at the scale $\Lambda$. Thus the consistency requires $M_{GUT} \lesssim 10^{17}\,$GeV with $N(M_{GUT}) \simeq 10^3$ in the
model under investigation, and thus clearly offers great hope for the observation of proton decay. This is actually a quite
general feature of supersymmetric grand unification in spite of large threshold effects \cite{jaigia}.

\section{Summary and Outlook}

We have shown that Type II see-saw dominance in the neutrino mass matrix can be achieved in $SO(10)$ models without adding ad hoc fields for this purpose. The theory where the symmetry breaking is achieved by representations ${\bf 54}$ and ${\bf 45}$, although it requires a complete Yukawa sector, is the one where perturbativity of the gauge couplings near and above $M_{GUT}$ can most easily be ensured.

We find two different possibilities of fine-tuning the mass of the Type II triplet together with some fields that cancel out their RGE contributions. The first one is made of a complete $SU(5)$  {\bf 15}, and the triplet can be as light as $10^{14}$ GeV. It can be a messenger of supersymmetry breaking following the lines of \cite{Joaquim:2006uz}, preserving basically all their predictions. The second one is a combination not forming a complete representation of an $SO(10)$ subgroup, which includes a pair of  doublets similar to the MSSM Higgs. In this case the triplet scale can be one order of magnitude lower, but the role of these light fields as supersymmetry breaking messengers is not completely clear. Namely, the  SM scale doublet would be a natural messenger too, and a careful study is needed before one can know whether this works or not.
While in the MSSM this cannot work since the stop becomes tachyonic, here there is in principle enough freedom to be realistic, an interesting question that  however requires a detailed analysis beyond the scope of this paper. .

The BSV mechanism can be realized for these light triplets in both cases.  The ${\bf 120}$ Yukawa coupling is shown to preserve the connection between $b-\tau$ unification and the quark and  neutrino mixing angles. This is very important for it illustrates nicely the possibility of the co-existence of small quark and large lepton mixing angles.

Admittedly, the theory becomes baroque and we do not propose it as the
minimal realistic supersymmetric grand unified theory. We do not claim that is the road to follow, but we take it rather as an illustration of the price one needs to pay in order to have a transparent picture of Type II seesaw in the context of renormalizable supersymmetric SO(10). Perhaps the most disappointing
aspect of this and similar theories is the impossibility of making predictions for proton decay before one knows the masses and mixings of the
superpartners. We still believe that such studies as ours are useful for they show that a great deal of information can be obtained on the
heavy particle spectra in spite of the complexity of the representations involved. We hope that in this sense this work can be of use to
practitioners in the field.

\section{Acknowledgments}

The authors wish to thank C.S. Aulakh for valuable comments and discussions. 
The work of  A.R.is supported by a grant from the "Plan II de Formaci\'on e Intercambio Cient\'ifico de la Universidad de Los Andes''. 
The work of G.S. was partially supported by the EU FP6 Marie Curie Research and Training Network "UniverseNet" (MRTN-CT-2006-035863). AM wishes to thank ICTP for hospitality  during the early stages of this work.

\section*{Appendix: Mass spectrum of $SO(10)$ theory with ${\bf 54}$ and ${\bf 45}$}

After symmetry breaking, the  $SO(10)$ representations decompose into 19 different states, and for most of them there is more than one particle, resulting in mass matrices mixing the contributions form the different  $SO(10)$ representations. The  mass matrices were calculated directly from the superpotential, written explicitly as

\begin{eqnarray}
W& =&\frac{m_1}{4} A_{ij} A_{ij} +  \frac{m_2}{2} H_{i} H_{i} +  \frac{m_3}{12} C_{ijk} C_{ijk}  + \frac{m_4}{4} S_{ij} S_{ij}
+ \frac{m_5}{5!}  \Sigma_{ijklm} \overline\Sigma_{ijklm}
\nonumber \\
&+& \lambda_1  A_{ij}A_{jk}S_{ki} + \frac{\lambda_2}{2} H_i H_j S_{ij} + \frac{\lambda_3}{2} C_{ijk} C_{ijl} S_{lk} + \frac{\lambda_4}{3} S_{ij}S_{jk}S_{ki}  \nonumber \\
&+&  \frac{\lambda_5}{4!} \Sigma_{ijklm} \Sigma_{ijklp} S_{mp}  +  \frac{\lambda_6}{4!} \overline\Sigma_{ijklm} \overline\Sigma_{ijklp} S_{mp}  +  \frac{\eta}{4!}  \Sigma_{ijklm} \overline\Sigma_{ijklp} A_{mp}   \nonumber \\
&+&  \frac{\beta}{6}  A_{ij} C_{klm}  \Sigma_{ijklm}  +  \frac{\bar \beta}{6}  A_{ij} C_{klm} \overline\Sigma_{ijklm}
+  \frac{\alpha}{2}  A_{ij}  H_k C_{ijk}
\end{eqnarray}
where
$$
S= {\bf 54}  , A= {\bf 45} , C= {\bf 120} , H= {\bf 10}, \Sigma = {\bf 126} , \overline\Sigma = \overline{\bf 126}
$$
We first identify all the states in the  $SO(10)$ representations, using the same conventions as in \cite{partone}.  Then the full superpotential is written in terms of all states and vevs, and for each pair of states $(\psi_i, \bar\psi_j) $ the fermion mass matrices are found by
 \beq
 M_{ij} = \left[ \frac{\partial^2 W }{\partial\bar\psi_i \partial\psi_j } \right]_{\phi_i=0}
 \eeq
We give  here the results, written in terms of the vevs in (\ref{thevevs}). States are identified by their $SU(3)_C,SU(2)_L,SU(2)_R$ quantum numbers and their $Y/2$. The $T_{3L}, T_{3_R}$ values are given as superscripts when necessary ($\pm $being a shorthand for $\pm 1/2$).

\begin{enumerate}
\item 
$({\bf 54}_{\overline{6}11}, {\bf 126}_{\overline{6}13^{+}}) $ , $ Y/2=\pm 2/3 $
$$
M_1=\left[\begin {array}{cc} m_{{4}}-8\,\lambda_{{4}}s&2\,\lambda_{{5}}
\sigma\,\sqrt {2}\\ \noalign{\medskip}2\,\lambda_{{6}} \bar\sigma\,\sqrt {2
}&m_{{5}}-\eta\,(a - 2b)\end {array}
 \right]
$$
\item $
({\bf 54}_{811}, {\bf 45}_{811}) $, $ Y/2=0$
 $$
M_2=\left[\begin {array}{cc} m_{{4}}-8\,\lambda_{{4}}s&4\,\lambda_{{1}}a
\\ \noalign{\medskip}-4\,\lambda_{{1}}a&m_{{1}}+8\,\lambda_{{1}}s
\end {array}
 \right]
$$
\item $
({\bf 54}_{322^{+}}, {\bf 45}_{322^{+}},  {\bf 120}_{322^{-}} , {\bf 126}_{322^{-}}, {\bf \overline{126}}_{322^{-}} )$, $  Y/2=\pm 1/6$
\begin{displaymath}
M_3=\left[\begin {array}{ccccc} 2\,\lambda_{{4}}s+m_{{4}}& 2\,\lambda_{{1}}(a + b)&0&0&2\,\lambda_{{6}}\bar\sigma\,\sqrt {2}
\\ \noalign{\medskip} -2\,\lambda_{{1}}(a+b)&-2\,\lambda_
{{1}}s+m_{{1}}&2\,\bar\beta \bar \sigma&\eta\,\bar\sigma\,\sqrt {2}&0
\\ \noalign{\medskip}0&2\,\beta \sigma&-2\,\lambda_{{3}}s+m_{{3}}
&\beta \sqrt {2} (a-b)&\bar\beta \sqrt {2} (a+b) \\ \noalign{\medskip}0&-\eta\, \sigma\,\sqrt {2}&-
\bar\beta \sqrt {2} (a-b) &\,\eta\,(2a -b) +m_{{5}
}&-10\,\lambda_{{6}}s\\ \noalign{\medskip}2\,\lambda_{{5}}\sigma\,
\sqrt {2}&0&-\beta \sqrt {2} (a+b)&-10\,\lambda_
{{5}}s&-\,\eta\,(2a -b) +m_{{5}}\end {array}
\right]
\end{displaymath}

\item $
({\bf 54}_{322^{-}}, {\bf 45}_{322^{-}})$, $Y/2=\pm 5/6$
 \begin{displaymath}
M_4=\left[\begin {array}{cc} m_{{4}} +  2\,\lambda_{{4}}s& 2\,\lambda_{{1}} (a - b)\\ \noalign{\medskip} -2\,\lambda_{
{1}}(a - b)& m_{{1}} -2\,\lambda_{{1}}s\end {array}
\right]
\end{displaymath}

\item $
({\bf 54}_{133^{0}}, {\bf 45}_{131})$, $Y/2=0$
 \begin{displaymath}
M_5=\left[\begin {array}{cc} m_{{4}}+12\,\lambda_{{4}}s&-4\,\lambda_{{1}
}b\\ \noalign{\medskip}4\,\lambda_{{1}}b&-12\,\lambda_{{1}}s+m_{{1}}
\end {array}
 \right]
\end{displaymath}

\item $
({\bf 54}_{133^{-}},{\bf 126}_{131}) $, $ Y/2= \pm 1$
\begin{displaymath}
M_6=\left[ \begin {array}{cc} m_{{4}}+12\,\lambda_{{4}}s&2\,\lambda_{{5}}
\sigma\,\sqrt {2}\\ \noalign{\medskip}2\,\lambda_{{6}} \bar\sigma\,\sqrt {2
}&m_{{5}} -3\,\eta\,a\end {array}
 \right]
\end{displaymath}

\item $
({\bf 54}_{111}, {\bf 45}_{111},  {\bf 45}_{113^{0}}, {\bf 126}_{\overline{1}13^{-}},{\bf \overline{126}}_{113^{+}})$,  $ Y/2=0$
 \begin{displaymath}
M_7= \left[ \begin {array}{ccccc} m_4 + 4\,\lambda_{{4}}s &-4 \sqrt{2}/\sqrt{5}\,\lambda_
{{1}}a&4\sqrt {3}/\sqrt {5}\,\lambda_{{1}}b &0&0
\\
\noalign{\medskip}-4 \sqrt{2}/\sqrt{5}\,\lambda_{{1}}a &- m_1-8\,
\lambda_{{1}}s&0&\eta\,\bar\sigma\,\sqrt {3}&\eta\,\sigma\,\sqrt {
3}\\
\noalign{\medskip}4\sqrt {3}/\sqrt {5}\,\lambda_{{1}}b &0&- m_1 +12\,
\lambda_{{1}}s &-\eta\,\bar\sigma\,\sqrt {2}&-\eta\,\sigma\,\sqrt {
2}\\
 \noalign{\medskip}0&\eta\,\bar\sigma\,\sqrt {3}&-\eta\,\bar\sigma\,\sqrt
{2}&m_{{5}}+ \eta(3\,a-2\,b)&0\\
 \noalign{\medskip}0&\eta\,\sigma
\,\sqrt {3}&-\eta\,\sigma\,\sqrt {2}&0& m_{{5}}+ \eta(3\,a-2\,b)
\end {array} \right]
\end{displaymath}

\item $
({\bf 45}_{311},  {\bf 120}_{313^{+}}, {\bf \overline{126}}_{313^{+}})$, $ Y/2=\pm 2/3$
\begin{displaymath}
M_8= \left[ \begin {array}{ccc} m_{{1}}+8\,\lambda_{{1}}s&2\,\beta
\sigma&-\eta\,\sigma\,\sqrt {2}\\ \noalign{\medskip}2\,\bar\beta
\bar\sigma&m_{{3}}+ 8\,\lambda_{{3}}s &-2\,\bar\beta a\sqrt {2}
\\ \noalign{\medskip}\eta\,\bar\sigma\,\sqrt {2}&2\,\beta a\sqrt {2}&m_{{5}} +
\eta\, (a -2 b) \end {array} \right]
\end{displaymath}

\item $
({\bf 45}_{113^{+}}, {\bf 120}_{\overline{1}11} , {\bf 126}_{\overline{1}13^{0}})$, $Y/2=\pm 1$

\begin{displaymath}
M_9=\left[ \begin {array}{ccc} m_{{1}} -12\,\lambda_{{1}}s &2\,\beta
\sigma&-\eta\,\sigma\,\sqrt {2}\\
 \noalign{\medskip}2\,\bar\beta
\bar\sigma&m_{{3}}-12\,\lambda_{{3}}s&2\,\bar\beta b\sqrt {2}
\\
\noalign{\medskip}\eta\,\bar\sigma\,\sqrt {2}&-2\,\beta b\sqrt {2}
& m_5 + 3\,\eta\,a \end {array} \right]
\end{displaymath}


 \item $({\bf \overline{126}}_{\overline{3}11},{\bf 126}_{\overline{3}11},{\bf 126}_{\overline{3}13^{0}},{\bf 120}_{\overline{3}11},  {\bf 120}_{\overline{3}13^{0}},{\bf 10}_{\overline{3}11})$, $Y/2=(+1/3,-1/3)$
\begin{displaymath}
M_{10}=\left[ \begin {array}{cccccc} m_{{5}}-\eta\,a&20\,\lambda_{{5}}s&0&2
\,\beta b&-2\,\beta a&0\\ \noalign{\medskip}20\,\lambda_{{6}
}s&m_{{5}}+\eta\,a&0&2\,\bar\beta b&2\,\bar\beta a&0
\\ \noalign{\medskip}0&0&m_{{5}}+\eta\,a&-2\,\bar\beta a\sqrt {2}&-2
\,\bar\beta b\sqrt {2}&0\\ \noalign{\medskip}-2\,\bar\beta b&-2\,
\beta b&2\,\beta a\sqrt {2}& m_{{3}}+ 8\,\lambda_{{3}}s&0&-
\alpha\,\sqrt {2}b\\ \noalign{\medskip}2\,\bar\beta a&-2\,\beta
a&2\,\beta b\sqrt {2}&0&m_{{3}}-12\,\lambda_{{3}}s&-\alpha\,
\sqrt {2}a\\ \noalign{\medskip}0&0&0&\alpha\,\sqrt {2}b&\alpha\,\sqrt
{2}a&m_{{2}}-2\,\lambda_{{2}}s\end {array} \right]
\end{displaymath}

\item $
( {\bf 120}_{\overline{3}31}, {\bf \overline{126}}_{\overline{3}31})$, $Y/2=\pm 1/3$
\begin{displaymath}
M_{11}= \left[ \begin {array}{cc} m_{{3}} + 8\,\lambda_{{3}}s&-2\,\bar\beta a
\sqrt {2}\\ \noalign{\medskip}2\,\beta a\sqrt {2}&m_{{5}}-\eta\,a
\end {array} \right]
\end{displaymath}

\item $
({\bf 120}_{611},  {\bf \overline{126}}_{613^{0}})$ , $ Y/2=\pm 1/3$
\begin{displaymath}
M_{12}= \left[ \begin {array}{cc} m_{{3}}-12\,\lambda_{{3}}s&2\,\bar\beta b
\sqrt {2}\\ \noalign{\medskip}-2\,\beta b\sqrt {2}&m_{{5}}-\eta\,
a\end {array} \right]
\end{displaymath}

\item $
({\bf 126}_{631})$,$  Y/2=\pm 1/3$
\begin{displaymath}
M_{13}=\left[ \begin {array}{c} { m_5}+{\eta}\,{a}\end {array} \right]
\end{displaymath}

\item $
({\bf 126}_{\overline{1}13^{+}})$, $ Y/2=\pm 2$
\begin{displaymath}
M_{14}=\left[\begin {array}{c} { m_5}+{\eta}\,(3{a}+2\,{b})\end {array} \right]
\end{displaymath}

\item $({\bf 120}_{313^{-}}, {\bf \overline{126}}_{313^{-}})$, $ Y/2=\pm 4/3$
\begin{displaymath}
M_{15}= \left[ \begin {array}{cc} m_{{3}}+ 8\,\lambda_{{3}}s&-2\,\bar\beta a
\sqrt {2}\\ \noalign{\medskip}2\,\beta a\sqrt {2}&m_{{5}}+\eta
\,(a+2b) \end {array} \right]
\end{displaymath}

\item $({\bf 126}_{\overline{6}13^{-}})$,$  Y/2=\pm 4/3$
 \begin{displaymath}
M_{16}=\left[ \begin {array}{c} {m_5}-{\eta}\,({a}+2\,{b})\end {array} \right]
\end{displaymath}


\item $
(\overline{{\bf 126}}_{122^{-}},{\bf 126}_{122^{-}},   {\bf 120}_{1'22^{-}},  {\bf 120}_{122^{-}},{\bf 10}_{122^{-}}) $, $ Y/2=\pm 1/2$
 \begin{displaymath}
M_{17}=\left[ \begin {array}{ccccc} m_{{5}}-\eta\,b&-10\,\lambda_{{5}}s&\,
\beta \sqrt {2}(2 a-b) &-\beta a\sqrt {2}
\sqrt {3}&0\\
 \noalign{\medskip}-10\,\lambda_{{6}}s&m_{{5}}+\eta\,b&
\,\bar\beta \sqrt {2} (2 a+b) &\bar\beta a\sqrt {2}
\sqrt {3}&0\\
 \noalign{\medskip}-\,\bar\beta \sqrt {2}(2 a-b) &-\,\beta \sqrt {2} (2 a+b) & m_{{3}}-2\,\lambda_
{{3}}s&0&-\alpha\,a\sqrt {3}\\
 \noalign{\medskip}\bar\beta a
\sqrt {2}\sqrt {3}&-\beta a\sqrt {2}\sqrt {3}&0&m_{{3}}+18\,
\lambda_{{3}}s&-\alpha\,b\\ \noalign{\medskip}0&0&\alpha\,a\sqrt {3}&
\alpha\,b&m_{{2}}+3\,\lambda_{{2}}s\end {array} \right]
\end{displaymath}

\item $
({\bf 120}_{\overline{3}22^{-}}, {\bf \overline{126}}_{\overline{3}22^{-}},{\bf 126}_{\overline{3}22^{-}})$, $ Y/2=\pm 7/6$
\begin{displaymath}
M_{18}= \left[ \begin {array}{ccc}m_{{3}} -2\,\lambda_{{3}}s&-\bar\beta
\sqrt {2} (a-b) &-\beta \sqrt {2} (a+b) \\ \noalign{\medskip}\beta \sqrt {2} (a-b) &m_{{5}} -\eta\,(2a+b)&-10\,\lambda_{{5}}s
\\ \noalign{\medskip}\bar\beta \sqrt {2} (a+b) &-10\,
\lambda_{{6}}s&m_{{5}} +\eta\,(2a+b)\end {array} \right]
\end{displaymath}

\item $\quad
({\bf 120}_{822^{+}},{\bf \overline{126}}_{822^{+}}, {\bf 126}_{822^{+}})$, $Y/2=\pm 1/2$
\begin{displaymath}
M_{19}= \left[ \begin {array}{ccc} m_{{3}} -2\,\lambda_{{3}}s&-\bar\beta
\sqrt {2} (a-b) &-\beta \sqrt {2} (a+b) \\ \noalign{\medskip}\beta \sqrt {2} (a-b)
&m_{{5}}+\eta\,b&-10\,\lambda_{{5}}s\\ \noalign{\medskip}\bar \beta \sqrt {2} (a+b) &-10\,\lambda_{{6}}s&m_{{5}}-\eta\,b
\end {array} \right]
\end{displaymath}

\end{enumerate}

\end{document}